\documentclass{INTERSPEECH2023}
\usepackage{bm}
\usepackage{graphicx}
\usepackage{float}
\usepackage{ascmac} 
\usepackage{fancybox}
\usepackage{float}
\usepackage{tabularx}
\usepackage{booktabs}
\usepackage{amssymb}
\usepackage{multirow}
\usepackage{cite}
\usepackage{multirow}
\usepackage[table,xcdraw]{xcolor}
\usepackage[normalem]{ulem}
\useunder{\uline}{\ul}{}
\usepackage{url}
\usepackage{pifont}
\usepackage{comment}

\AtBeginDocument{
  \abovedisplayskip     =.8\abovedisplayskip
  \abovedisplayshortskip=.8\abovedisplayshortskip
  \belowdisplayskip     =.8\belowdisplayskip
  \belowdisplayshortskip=.8\belowdisplayshortskip}

\interspeechcameraready

\title{Knowledge Distillation for Neural Transducer-based Target-Speaker ASR: Exploiting Parallel Mixture/Single-Talker Speech Data}
\name{Takafumi Moriya, Hiroshi Sato, Tsubasa Ochiai, Marc Delcroix, Takanori Ashihara, \\ Kohei Matsuura, Tomohiro Tanaka, Ryo Masumura, Atsunori Ogawa, Taichi Asami}
\address{\begin{tabular}{c} NTT Corporation, Japan \end{tabular}}
\email{takafumi.moriya.nd@hco.ntt.co.jp}

\ninept

\begin{document}

\maketitle
\begin{abstract} 
Neural transducer (RNNT)-based target-speaker speech recognition (TS-RNNT) directly transcribes a target speaker's voice from a multi-talker mixture. It is a promising approach for streaming applications because it does not incur the extra computation costs of a target speech extraction frontend, which is a critical barrier to quick response. TS-RNNT is trained end-to-end given the input speech (i.e., mixtures and enrollment speech) and reference transcriptions. The training mixtures are generally simulated by mixing single-talker signals, but conventional TS-RNNT training does not utilize single-speaker signals. This paper proposes using knowledge distillation (KD) to exploit the parallel mixture/single-talker speech data. Our proposed KD scheme uses an RNNT system pretrained with the target single-talker speech input to generate pseudo labels for the TS-RNNT training. Experimental results show that TS-RNNT systems trained with the proposed KD scheme outperform a baseline TS-RNNT.
\end{abstract}

\section{Introduction}
Although recent advances in end-to-end automatic speech recognition (ASR) frameworks have boosted ASR performance in the single-talker case~\cite{anmol2020conformer,Sainath2021AnES,chen2021lcconformer,kurata2020kdrnnt,panchapagesan2021kdrnnt},
it remains difficult to recognize multi-talker speech if multiple voices are overlapped~\cite{barker2018fifth}. 
Target-speaker ASR (TS-ASR)~\cite{moriya2022tsrnnt,moriya2023tsasrad}, which recognizes only the target speaker's speech from a mixture, is a promising technology for developing user-dependent voice interactive systems including smart speakers, smartphones, smartwatches, etc. 
In this paper, we focus on improving the TS-ASR performance.

One conventional approach to realizing TS-ASR is to use a cascade of a target speech extraction (TSE) frontend with an ASR backend~\cite{vzmolikova2019speakerbeam,delcroix2018single,wangvoicefilter,sato2021SE,Sato2021switch,ShiZW2021jointmtasr}.
First, the TSE frontend extracts the target speaker's voice from the speech mixtures by utilizing a speaker embedding that characterizes the target speaker. This embedding is derived by processing the enrollment utterance of the target speaker in a speaker encoder module.
Then, the speech extracted by the TSE module is fed to the ASR module to obtain the target speaker's transcription.

In the training step of TSE, signal-level loss is computed using the estimated target speaker's speech and the reference clean speech of the target speaker. 
Thus, TSE training needs parallel data consisting of mixtures and the target speaker's reference/enrollment speech. 
In the inference step, the cascade approach improves the TS-ASR performance. However, the extra computation costs of the TSE module are a critical barrier to quick response, especially for streaming operation. 

Recently,~\cite{moriya2022tsrnnt,moriya2023tsasrad} proposed an integrated TS-ASR framework that incorporates the TSE essence (i.e., conditioning on speaker embedding) into a neural transducer (RNNT)~\cite{Graves2012} model, called TS-RNNT. 
TS-RNNT can identify and transcribe the target speaker's speech while ignoring the interfering speakers' voices, by conditioning an RNNT model on the target speaker embedding. The computation cost is equivalent to that of a conventional RNNT model. 

TS-RNNT training requires the input speech signals, i.e., mixtures and target speaker's enrollment utterances, and the target speaker's transcription. 
Although TS-RNNT training does not require single-talker speech, such clean source data is generally used to generate the simulated mixtures. 
Thus the use of single-talker speech is logical for TS-RNNT training. 
However, prior works~\cite{moriya2022tsrnnt,moriya2023tsasrad} did not directly utilize the plain single-talker speech for TS-RNNT training. 
We expect that we could improve TS-ASR performance further if we could utilize the parallel mixture/target speech for TS-RNNT training. 

The TS-RNNT system is an end-to-end TS-ASR system that directly performs recognition without explicitly extracting the target speech signal. 
Therefore, it is not straightforward to utilize parallel mixture and single-talker speech data because it is not possible to define a signal-level loss term as is used in TSE. 
As an alternative, in this paper, we propose a knowledge distillation (KD)-based training framework to exploit the parallel mixture/single-talker speech data. 

We generate pseudo labels (posteriors) by processing the single-talker speech with a teacher RNNT model trained on single-talker data. We then train the student TS-RNNT model using a multi-task loss consisting of the standard RNNT loss and a KD loss. 
The KD loss is the cross-entropy loss between the posteriors computed with the TS-RNNT (i.e., student) and the pseudo labels. 
The teacher's guidance in the KD loss may help the student learn posteriors that are more robust to the interference speakers. Thus, we expect that the TS-RNNT trained with the proposed KD will improve the TS-ASR performance.

We conduct experiments on parallel mixture/single-talker speech data, to compare our TS-RNNT proposal, trained with the proposed KD, called TS-RNNT+KD, with conventional TS-RNNT in both offline and streaming modes. 
Experiments show that TS-RNNT+KD outperforms the variant without KD, especially for streaming TS-ASR. 

\vspace{-0.15cm}
\section{Related Work}
\vspace{-0.05cm}
While the KD framework is generally used to relieve performance degradation in model compression~\cite {panchapagesan2021kdrnnt,knowledge2014hinton,jinyu2014kd}, some previous studies apply it for other purposes~\cite{kdda2017asami,watanabe2017kd,Fukuda2017,wenjie2018kdclose2far,tian2018kdmtasr,zhang19kde2emtasr,zhong2019e2ekd,moriya2020daw,wangyou2020kdmtasr,kurata2020kdrnnt,moriya2020sdaed,Kojima2021KnowledgeDF}. 
In particular, the KD frameworks  of~\cite{tian2018kdmtasr,zhang19kde2emtasr,wangyou2020kdmtasr} have been used for multi-talker ASR systems, where a single-talker ASR system with the target speaker's speech input is used to generate pseudo-labels. 
These works dealt with ``separation'' based systems that recognize the speech of all speakers in a mixture and identification of a target speaker is not considered. 
These frameworks require Permutation Invariant Training loss, making the training more complex due to the multiple decoders within the models as well as RNNT-based multi-talker ASR systems~\cite{SklyarPL21PIT,Lu2021surt}. 
They explored KD for neural network-Hidden Markov Model hybrid~\cite{tian2018kdmtasr}, and attentional encoder-decoder~\cite{zhang19kde2emtasr,wangyou2020kdmtasr} multi-talker ASR-based systems, and then only for offline settings.

In this work, we use KD for TS-RNNT, which simplifies the training as there is only a single target speaker. 
Thus TS-RNNT can naturally avoid output-speaker ambiguity and identify the target speaker. 
We use the approach developed for more recent RNNT systems, which allows the exploration of KD for streaming models.


\begin{figure}[t]
 \begin{center}
    \includegraphics[width=8.2cm]{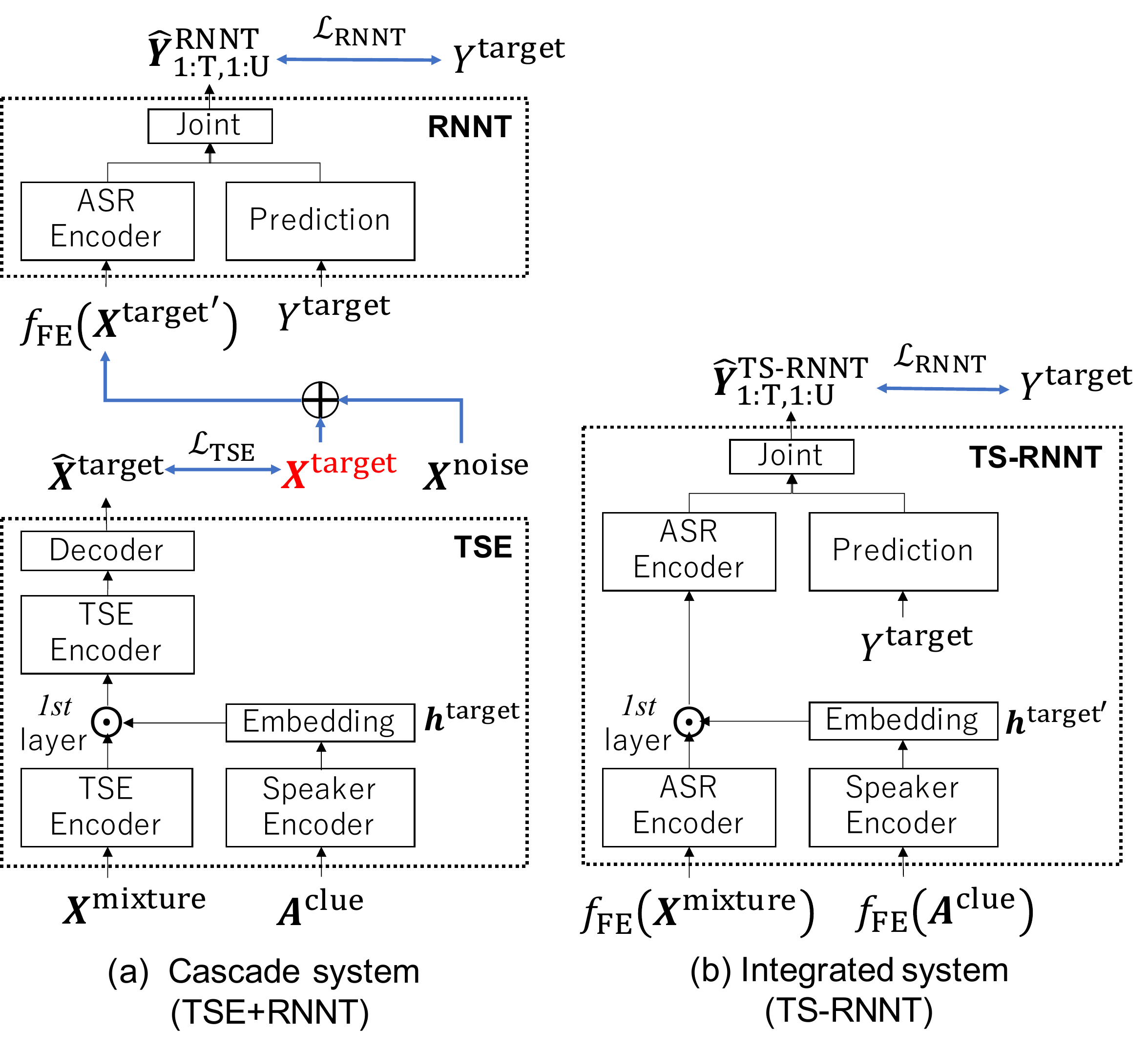}
 \end{center}
\vspace{-0.65cm}
 \caption{Overview of (a) cascaded and (b) integrated TS-ASR systems. $\bm{X}^{\text{target}}$ is only used to train the cascaded system.}
 \vspace{-0.55cm}
 \label{fig:framework}
\end{figure}

\begin{figure*}[t]
 \begin{center}
    \includegraphics[width=16.6cm]{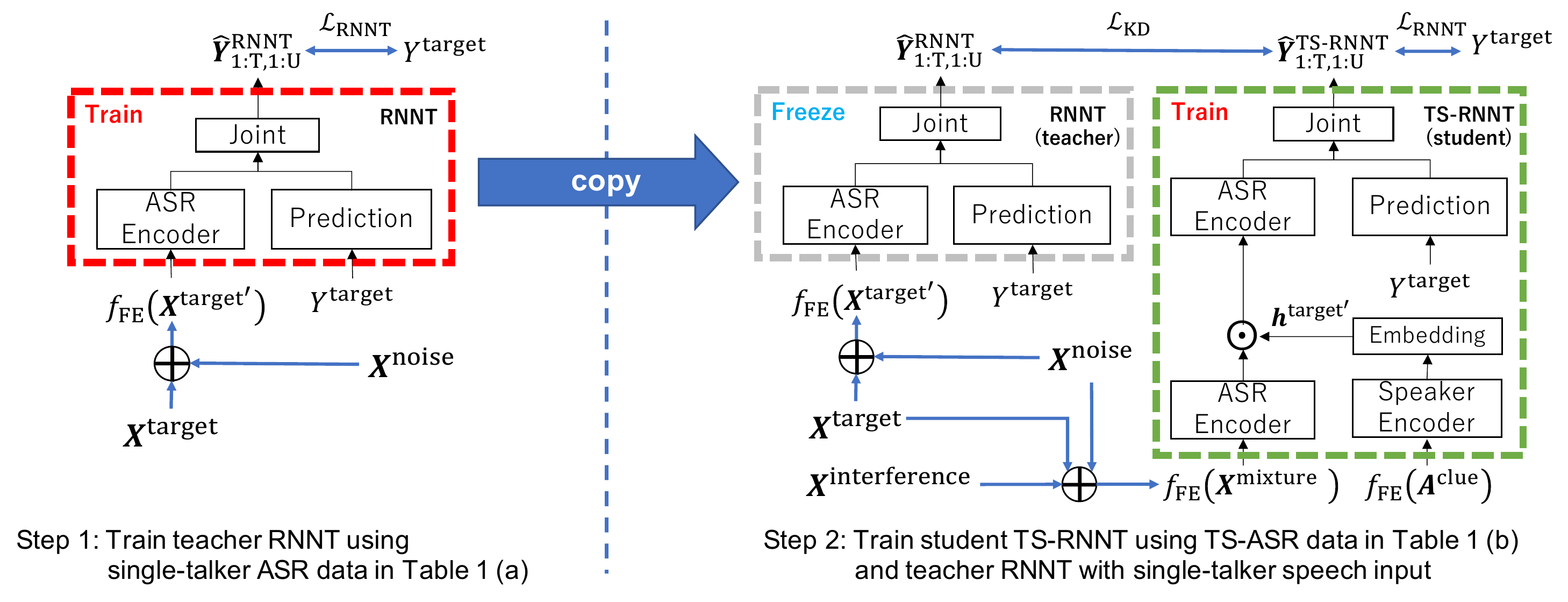}
 \end{center}
\vspace{-0.7cm}
 \caption{Schematic diagram of TS-RNNT+KD for TS-RNNT. The training process consists of two steps: 1) pretraining the teacher RNNT using single-talker ASR data with $\mathcal{L}_{\text{RNNT}}$ as the loss, and 2) training the student TS-RNNT using the mixture and target speaker's transcription pair data (TS-ASR data) and the pre-trained RNNT outputs with multi-task loss $\mathcal{L}_{\text{RNNT+KD}}$ combining $\mathcal{L}_{\text{RNNT}}$ and $\mathcal{L}_{\text{KD}}$.}
 \label{fig:tsasr_kd}
\end{figure*}

\vspace{-0.2cm}
\section{Methods}
In this section, we first explain the cascaded TS-ASR, i.e., TSE+RNNT, and the integrated TS-ASR, i.e., TS-RNNT, which are the foundations of this work. 
Then we introduce our proposed approach, i.e., TS-RNNT+KD. 

Let $\bm{X}^{\text{mixture}} = \left[ x^{\text{mixture}}_{1}, ..., x^{\text{mixture}}_{T^{\prime}} \right] \in \mathcal{R}^{T^{\prime}}$ be the single microphone input speech mixture of duration $T^{\prime}$; it includes the target speech $\bm{X}^{\text{target}}$, interfering speakers' voices $\bm{X}^{\text{interference}}$, and background noise $\bm{X}^{\text{noise}}$. 
$Y^{\text{target}} = \left[ y^{\text{target}}_{1}, ..., y^{\text{target}}_{U} \right] \in \mathcal{R}^{U}$ is the sequence of tokens of length $U$ associated with the utterance spoken by the target speaker, where $y^{\text{target}}_u \in \{1, ..., K\}$. 
$K$ is the number of tokens in the vocabulary. 

\vspace{-0.2cm}
\subsection{Cascaded TS-ASR (TSE+RNNT)}
\label{ssec:cascade}
\vspace{-0.1cm}
Figure~\ref{fig:framework}~(a) is a diagram of the cascaded TS-ASR system, which is composed of the TSE and ASR modules described below.

\vspace{-0.25cm}
\subsubsection{TSE frontend}
\label{sssec:se}
\vspace{-0.1cm}
In this work, we adopt the time-domain TSE module~\cite{sato2021SE,Sato2021switch} that extracts $\hat{\bm{X}}^{\text{target}}$ from $\bm{X}^{\text{mixture}}$ using enrollment speech $\bm{A}^{\text{clue}}$. 
First, we extract the embedding of the target-speaker, $\bm{h}^{\text{target}}$, from $\bm{A}^{\text{clue}}$ using speaker encoder $f^{\text{Spk-Enc}}(\cdot)$, which consists of a multi-layer neural network followed by a linear layer and a time-average pooling layer. 
Then, we use a speech extraction network $f^{\text{SE}}(\cdot)$, which consists of TSE encoder and decoder, to extract the target speech given the embedding, $\bm{h}^{\text{target}}$. 
The above operations, which yield the estimated target speaker's speech signal $\hat{\bm{X}}^{\text{target}}$, are defined as follows:
\begin{eqnarray}
\bm{h}^{\text{target}} &=& f^{\text{Spk-Enc}} (\bm{A}^{\text{clue}}; \theta^{\text{Spk-Enc}}), \\
\hat{\bm{X}}^{\text{target}} &=& f^{\text{SE}} (\bm{X}^{\text{mixture}}, \bm{h}^{\text{target}}; \theta^{\text{SE}}). 
\label{eq:se}
\end{eqnarray}
In this work, we utilize the Hadamard product between $\bm{h}^{\text{target}}$ and the first encoder layer output of $f^{\text{SE}} (\cdot)$. 
The parameters $\theta^{\text{TSE}} \triangleq [\theta^{\text{Spk-Enc}}, \theta^{\text{SE}}]$ are jointly optimized with the scale-invariant source-to-noise ratio (SI-SNR) loss $\mathcal{L}_{\text{TSE}}$~\cite{luo2019conv} using $\hat{\bm{X}}^{\text{target}}$ and reference $\bm{X}^{\text{target}}$.

\begin{table*}[th]
\vspace{-0.2cm}
\centering
\caption{Data generation setup. The number of utterances in (a) is double of (b) due to the single-speaker case of (b).}
 \vspace{-0.3cm}
\label{tab:setup}
\scalebox{0.89}[0.89]{
\begin{tabular}{lllcccc}
\hline
 & \multicolumn{1}{c}{dataset} & \multicolumn{1}{c}{mixture type} & \begin{tabular}[c]{@{}c@{}}SIR\\ {[}dB{]}\end{tabular} & \begin{tabular}[c]{@{}c@{}}SNR\\ {[}dB{]}\end{tabular} & \begin{tabular}[c]{@{}c@{}}\#speakers\\ (train set / dev set)\end{tabular} & \begin{tabular}[c]{@{}c@{}}\#mixtures or utterances\\ (train set / dev set)\end{tabular} \\ \hline
(a) & training data for ASR backend (RNNT) & 1 speaker and noise & - & 0 - 20 & 3054 / 160 & 400,000 / 10,000 \\
(b) & training data for TSE and TS-RNNT & 2 speakers and noise & -5 - 5 & 0 - 20 & 3054 / 160 & 200,000 / 5,000 \\
(c) & evaluation data & 2 speakers and noise & -5 - 5 & 0, 5, 10, 15, 20 & 30 & $6,000 \times 5 = 30,000$ \\ \hline
\end{tabular}
}
\vspace{-0.4cm}
\end{table*}

\vspace{-0.2cm}
\subsubsection{ASR backend (RNNT)}
\label{sssec:rnnt}
\vspace{-0.1cm}
We use an RNNT-based ASR model~\cite{Graves2012}, which can perform streaming ASR, for the backend module. 
RNNT learns the mapping between sequences of different lengths. 
First, in the training step, $\bm{X}^{\text{target}^{\prime}}$ consisting of $\bm{X}^{\text{target}}$ with background noise $\bm{X}^{\text{noise}}$ to enhance system robustness is transformed into acoustic features using feature extractor $f^{\text{FE}}(\cdot)$. 
Then, the features are encoded into a length $T$ sequence, $\bm{H}^{\text{ASR}}$, via ASR encoder network $f^{\text{ASR-Enc}}(\cdot)$.
Next, $Y^{\text{target}}$ is also encoded into
$\bm{H}^{\text{Pred}}$
via a prediction network $f^{\text{Pred}}(\cdot)$.
$\bm{H}^{\text{ASR}}$ and $\bm{H}^{\text{Pred}}$ are fed to a joint network, $f^{\text{Joint}}(\cdot)$, to compute the token posterior probabilities, $\hat{\bm{Y}}^{\text{RNNT}}_{1:T,1:U} \in \mathcal{R}^{T \times U \times K}$.
The above operations can be formularized as follows:
\begin{eqnarray}
\bm{H}^{\text{ASR}} &=& f^{\text{ASR-Enc}} (f^{\text{FE}}(\bm{X}^{\text{target}^{\prime}}); \theta^{\text{ASR-Enc}}), \\
\bm{H}^{\text{Pred}} &=& f^{\text{Pred}} (Y^{\text{target}}; \theta^{\text{Pred}}), \\
\hat{\bm{Y}}^{\text{RNNT}}_{1:T,1:U} &=& \text{Softmax} \left(f^{\text{Joint}} (\bm{H}^{\text{ASR}}, \bm{H}^{\text{Pred}}; \theta^{\text{Joint}}) \right),
\label{eq:rnnt}
\end{eqnarray}
where $\text{Softmax}(\cdot)$ performs a softmax operation. 
All the parameters $\theta^{\text{RNNT}} \triangleq [\theta^{\text{ASR-Enc}}, \theta^{\text{Pred}}, \theta^{\text{Joint}}]$ are optimized with RNNT loss $\mathcal{L}_{\text{RNNT}}$ using $\hat{\bm{Y}}^{\text{RNNT}}_{1:T,1:U}$ and $Y^{\text{target}}$. 

For training, we use single-talker speech $\bm{X}^{\text{target}^{\prime}}$ consisting of $\bm{X}^{\text{target}}$ mixed with $\bm{X}^{\text{noise}}$ and the transcript $Y^{\text{target}}$.
Note that we do not perform joint training of the TSE and RNNT models. 
In the decoding step, we use $\hat{\bm{X}}^{\text{target}}$ extracted from the TSE.

\vspace{-0.25cm}
\subsection{Integrated TS-ASR with RNNT (TS-RNNT)}
\label{ssec:tsrnnt}
\vspace{-0.05cm}
Figure~\ref{fig:framework} (b) is a schematic diagram of the TS-RNNT system. 
TS-RNNT deals with the target speaker embedding $\bm{h}^{\text{target}^{\prime}}$ to inform which speaker in the mixture to decode; it can directly transduce the mixture to yield the target speaker's transcription. 
TS-RNNT encoder $f^{\text{ASR-Enc}^{\prime}}(\cdot)$ receives $f^{\text{FE}}(\bm{X}^{\text{mixture}})$ and $\bm{h}^{\text{target}^{\prime}}$ extracted from $f^{\text{FE}}(\bm{A}^{\text{clue}})$ via $f^{\text{Spk-Enc}^{\prime}} (\cdot)$ of TS-RNNT, and outputs $\bm{H}^{\text{ASR}^{\prime}}$. 
These functions are defined as follows:
\begin{eqnarray}
\bm{h}^{\text{target}^{\prime}} &=& f^{\text{Spk-Enc}^{\prime}} (f^{\text{FE}} (\bm{A}^{\text{clue}}); \theta^{\text{Spk-Enc}^{\prime}}), \\
\bm{H}^{\text{ASR}^{\prime}} &=& f^{\text{ASR-Enc}^{\prime}} (f^{\text{FE}}(\bm{X}^{\text{mixture}}), \bm{h}^{\text{target}^{\prime}}; \theta^{\text{ASR-Enc}^{\prime}} ).  
\label{eq:tsrnnt}
\end{eqnarray}
$\bm{h}^{\text{target}^{\prime}}$ and the first layer ASR encoder output are multiplied as the Hadamard product within $f^{\text{ASR-Enc}^{\prime}} (\cdot)$. 
The prediction and joint networks are the same as in the standard RNNT of~\ref{sssec:rnnt}, and TS-RNNT also outputs the posterior probabilities in tensor format $\hat{\bm{Y}}^{\text{TS-RNNT}}_{1:T,1:U} \in \mathcal{R}^{T \times U \times K}$. 
All networks with parameters $\theta^{\text{TS-RNNT}} \triangleq [\theta^{\text{Spk-Enc}^{\prime}}, \theta^{\text{ASR-Enc}^{\prime}}, \theta^{\text{Pred}}, \theta^{\text{Joint}}]$ are jointly optimized by using RNNT loss $\mathcal{L}_{\text{RNNT}}$.

For decoding, we enroll $\bm{h}^{\text{target}^{\prime}}$ extracted from $f^{\text{FE}}(\bm{A}^{\text{clue}})$ in advance. 
Then, the acoustic feature of the mixture signal, $f^{\text{FE}}(\bm{X}^{\text{mixture}})$, is directly input to the ASR encoder together with $\bm{h}^{\text{target}^{\prime}}$, and the decoder yields the TS-ASR results. 
Thanks to the removal of the TSE frontend module, TS-RNNT can perform TS-ASR faster than the cascaded system, while its computation cost equals that of basic RNNT. 
TS-RNNT training uses the target speakers' voices and the transcriptions, which are arguably easier to collect for real recordings; it does not need \textit{clean} target speakers' speech data as references for TSE module training. 

\vspace{-0.3cm}
\subsection{Proposed KD for TS-RNNT (TS-RNNT+KD)}
\label{ssec:tsasrkd}
\vspace{-0.1cm}
TS-RNNT does not require single-talker speech in the training step. 
However, speech mixtures are generally simulated using single-talker speech data. 
Therefore, target single-speaker speech $\bm{X}^{\text{target}}$ is naturally available for training.
In this paper, we exploit the parallel speech data, i.e., $\bm{X}^{\text{target}}$ and $\bm{X}^{\text{mixture}}$, using a KD framework. 

The schematic of our proposed KD approach of exploiting parallel speech data for TS-RNNT is illustrated in Figure~\ref{fig:tsasr_kd}. 
Our proposed KD for TS-RNNT training consists of two steps; 1) building a single-talker teacher model and 2) distilling knowledge from the pretrained single-talker RNNT with target speaker's speech input to a TS-RNNT model. 
The procedure is detailed as follows.

First, we train single-talker RNNT as a teacher model using $\bm{X}^{\text{target}^{\prime}}$, which consists of $\bm{X}^{\text{target}}$ mixed with $\bm{X}^{\text{noise}}$, and $Y^{\text{target}}$; the RNNT is optimized by $\mathcal{L}_{\text{RNNT}}$. 
We assume that the pretrained single-talker RNNT with single-talker speech input can better align the speech and the transcriptions, thus providing more reliable posteriors than those of TS-RNNT. 
Then, we train a student TS-RNNT using the mixture, and the pretrained teacher RNNT outputs $\hat{\bm{Y}}^{\text{RNNT}}_{1:T,1:U}$. This is achieved by feeding the single-talker speech $\bm{X}^{\text{target}^{\prime}}$ to the pretrained single-talker RNNT as the teacher. In this stage, we freeze the parameters of the single-talker RNNT model. 
We train the student model using multi-task loss $\mathcal{L}_{\text{RNNT+KD}} = \mathcal{L}_{\text{RNNT}} + \lambda \mathcal{L}_{\text{KD}}$, where $\mathcal{L}_{\text{RNNT}}$ is computed using $\hat{\bm{Y}}^{\text{TS-RNNT}}_{1:T,1:U}$ and $Y^{\text{target}}$ , and $\mathcal{L}_{\text{KD}}$ is the cross-entropy loss between $\hat{\bm{Y}}^{\text{RNNT}}_{1:T,1:U}$ and $\hat{\bm{Y}}^{\text{TS-RNNT}}_{1:T,1:U}$. $\lambda$ is a hyperparameter for balancing the losses between $\mathcal{L}_{\text{RNNT}}$ and $\mathcal{L}_{\text{KD}}$. KD loss is defined as follows:
\begin{equation}
  \mathcal{L}_{\text{KD}} =  - \sum_{t=1}^{T} \sum_{u=1}^{U} \sum_{k=1}^{K} \hat{y}^{\text{RNNT}}_{t,u,k} \ \log \ \hat{y}^{\text{TS-RNNT}}_{t,u,k},
  \label{eq:KD}
\end{equation}
where $\hat{y}^{\text{RNNT}}_{t,u,k}$ and $\hat{y}^{\text{TS-RNNT}}_{t,u,k}$ correspond to the $k$-th class probability of $\hat{\bm{Y}}^{\text{RNNT}}_{1:T,1:U}$ and $\hat{\bm{Y}}^{\text{TS-RNNT}}_{1:T,1:U}$ at the $t$-th time and $u$-th label steps, respectively. 
The KD loss $\mathcal{L}_{\text{KD}}$ replaces the signal-level loss on the TSE output of cascade signals when there is no explicit target speech signal estimation.

\begin{table*}[t]
\centering
\caption{Comparisons of baseline cascade, baseline integrated, and proposed TS-ASR systems. The system ID uses the following notations, ``B'',  ``P'', ``O'' ``S'', for baseline,  proposed, offline, and streaming, respectively. 
}
 \vspace{-0.25cm}
\label{tab:comp}
\scalebox{0.92}[0.92]{
\begin{tabular}{l|c||c|cccccc||c|cccccc}
\hline
\multirow{2}{*}{System} & \multirow{2}{*}{$\lambda$} & \multirow{2}{*}{ID} & \multicolumn{6}{c||}{CERs [\%] of offline systems on each SNR}                                                                     & \multirow{2}{*}{ID} & \multicolumn{6}{c}{CERs [\%] of streaming systems on each SNR}                                                                      \\
                        &                         &                     & 20dB         & 15dB         & 10dB          & 5dB           & \multicolumn{1}{c|}{0dB}           & Avg.          &                     & 20dB          & 15dB          & 10dB          & 5dB           & \multicolumn{1}{c|}{0dB}           & Avg.          \\ \hline
TSE+RNNT                & -                       & BO1                 & \textbf{7.9} & 8.8          & 11.1          & 18.3          & \multicolumn{1}{c|}{36.3}          & 16.5          & BS1                 & 16.8          & 18.6          & 23.8          & 35.6          & \multicolumn{1}{c|}{56.3}          & 30.2          \\ \hline
TS-RNNT                 & 0.0                     & BO2                 & 8.6          & 9.3          & 11.4          & 17.3          & \multicolumn{1}{c|}{32.7}          & 15.8          & BS2                 & 11.9          & 13.0          & 16.0          & 23.8          & \multicolumn{1}{c|}{41.2}          & 21.2          \\ \hline
\multirow{5}{*}{\begin{tabular}[c]{@{}l@{}}TS-RNNT+KD\\ (proposed)\end{tabular}}    & 1.0                     & PO1                 & 8.9          & 9.5          & 11.6          & 17.5          & \multicolumn{1}{c|}{34.2}          & 16.3          & PS1                 & 11.6          & 12.9          & 15.8          & 24.2          & \multicolumn{1}{c|}{42.2}          & 21.4          \\
                        & 0.5                     & PO2                 & 8.5          & 9.1          & 11.2          & 16.7          & \multicolumn{1}{c|}{33.0}          & 15.7          & PS2                 & 11.7          & 12.7          & 15.7          & 23.7          & \multicolumn{1}{c|}{41.4}          & 21.0          \\
                        & 0.1                     & PO3                 & 8.1          & \textbf{8.7} & \textbf{10.7} & \textbf{16.1} & \multicolumn{1}{c|}{\textbf{31.4}} & \textbf{15.0} & PS3                 & 11.0          & 12.2          & 15.0          & 22.4          & \multicolumn{1}{c|}{39.8}          & 20.1          \\
                        & 0.01                    & PO4                 & 8.5          & 9.1          & 11.0          & 17.0          & \multicolumn{1}{c|}{32.1}          & 15.5          & PS4                 & \textbf{10.2} & \textbf{11.3} & \textbf{14.2} & \textbf{21.7} & \multicolumn{1}{c|}{\textbf{38.7}} & \textbf{19.2} \\
                        & 0.001                   & PO5                 & 8.8          & 9.4          & 11.3          & 16.8          & \multicolumn{1}{c|}{31.7}          & 15.6          & PS5                 & 10.7          & 11.8          & 14.7          & 22.3          & \multicolumn{1}{c|}{39.2}          & 19.7          \\ \hline
\end{tabular}
}
\vspace{-0.4cm}
\end{table*}

\vspace{-0.25cm}
\section{Experiments}
\label{sec:result}
\vspace{-0.15cm}
\subsection{Data}
\label{ssec:data}
\vspace{-0.15cm}

For the evaluation, we used the Corpus of Spontaneous Japanese (CSJ)~\cite{maekawa2000} and the CHiME-3 corpus~\cite{chime3} to simulate the mixture. 
The details are shown in Table~\ref{tab:setup}. 
The mixture consists of speech signals taken from the CSJ corpus, with background noise taken from the CHiME-3 corpus at signal-to-noise ratio (SNR) between 0 and 20 dB. 
The training data contains samples totaling 800 hours. 
The overlap ratio of the mixtures in both training and evaluation datasets was about 89\% on average. 
The speakers between training, development, and evaluation datasets are different.
In this paper, we adopt 3262 characters for the ASR tasks. 
We evaluated performance in terms of character error rate (CER) due to the ambiguity of Japanese word boundaries. 

\vspace{-0.25cm}
\subsection{System configuration of TSE module}
\label{ssec:sesystem}
\vspace{-0.15cm}

We adopted a Conv-TasNet~\cite{luo2019conv}-based time-domain SpeakerBeam structure as a TSE frontend~\cite{delcroix2020improving,sato2021SE}. 
The details of the implementation are similar to those in~\cite{sato2021SE}.
In this work, we trained offline and streaming TSE models. 
All TSE models were trained with dataset (b) in Table~\ref{tab:setup}. 
Note that the convolution and global layer normalization of offline TSE were replaced with causal convolution and channel-wise layer normalization in the streaming one, respectively. 
The algorithmic latency of the streaming TSE model is 1.25ms, which is negligible in terms of ASR decoding latency. 
The source-to-distortion improvements in offline and streaming TSE models were 15.1dB and 11.1dB, respectively, which mirrors the tendency reported in prior separation studies~\cite{luo2019conv}. 

\vspace{-0.25cm}
\subsection{System configuration of ASR module}
\label{ssec:asrsystem}
\vspace{-0.1cm}

We used an 80-dimensional log Mel-filterbank as the input feature of ASR models. 
SpecAugment~\cite{specaugment} was applied to the feature during training. 
We investigated two versions of (TS-) RNNT. 
First, we tested with an offline system consisting of the same encoder architecture as Conformer (L)~\cite{anmol2020conformer} with a kernel size of 15. 
The ASR encoder contains two-layer 2D-convolutional neural networks (CNNs) followed by 17 Conformer blocks. The stride sizes of both max-pooling layers at each CNN layer were set to $2 \times 2$. 
 The prediction network had a 768-dimensional uni-directional long short-term memory layer.
The joint network consisted of a 640-dimensional feed-forward network and output the posterior probabilities. 
The speaker encoder for TS-RNNT had the same architecture as the ASR encoder, while the number of blocks was reduced from 17 to 6. 

We also performed a streaming experiment, 
and compared our proposed variant with a streaming system that used a similar configuration as the offline system except that the ASR encoder is replaced by the streaming Conformer encoder~\cite{chen2021lcconformer}. 
We adopted causal depthwise convolution and layer normalization for the streaming Conformer encoder that was trained with an attention mask strategy as in~\cite{chen2021lcconformer}.
The history and current chunk sizes of streaming Conformer were set to 68 and 60 frames, respectively. 
Thus the average latency was 330ms ($= 600\text{ms} / 2 + 30\text{ms}$). 

``RNNT'' and ``TS-RNNT'' were trained with the single-talker data (a) and the mixture data (b) in Table~\ref{tab:setup}, respectively. 
The parameters of offline Conformer model were randomly initialized. 
The streaming Conformer parameters were initialized with those of a trained offline Conformer. 
We used the Adam optimizer with 25k warmup for a total of 100 epochs, and all models were trained using RNNT loss. 
The minibatch size was set to 64 in all experiments. 
For decoding, we performed alignment-length synchronous decoding with beam width of 8~\cite{saon2020alsd}. 
We used the Kaldi and ESPnet toolkits~\cite{povey2011kaldi,espnet} for all implementation, data preprocessing, training, and evaluation processes.

\vspace{-0.25cm}
\subsection{Results}
\vspace{-0.1cm}
\subsubsection{Baseline TSE+RNNT vs. Baseline TS-RNNT} 
\vspace{-0.15cm}
First, we compare the baseline cascade system B*1 with baseline integrated systems B*2 for offline and streaming modes. 
The left and right blocks of Table~\ref{tab:comp} show the CERs under each SNR condition of the offline and streaming systems, respectively. 
Although the offline cascade system (BO1) contains the strong TSE model, the offline integrated system (BO2), i.e., TS-RNNT without any frontend module, achieves comparable or better CERs than BO1. 
We also compare the streaming cascade system (BS1) with the streaming integrated system (BS2). 
The CERs of BS2 are much better than those of BS1 under all conditions. 
The reason is that the performance of BS1 heavily depends on the quality of the TSE output. 
The TS-RNNT could avoid the performance degradation caused by using the streaming TSE module. 
Hereafter, the TS-RNNTs (BO2 and BS2) are regarded as the offline and streaming baseline TS-ASR systems, respectively. 

\vspace{-0.25cm}
\subsubsection{Baseline TS-RNNT vs. Proposed TS-RNNT+KD} 
\vspace{-0.15cm}
Next, we compare the baseline TS-RNNT systems (BO2 and BS2) with the proposed systems ``TS-RNNT+KD'' (PO* and PS*), which were trained with $\mathcal{L}_{\text{RNNT+KD}}$ described in~\ref{ssec:tsasrkd}. 
The CERs under each SNR condition are shown in Table~\ref{tab:comp}. 
The results of offline and streaming TS-RNNT+KD are denoted as ``PO*'' and ``PS*'', respectively. 

The offline experiment shows that the offline system PO3 achieved the best recognition performance among the TS-RNNT+KD system by tuning the hyperparameter $\lambda = 0.1$ on the development set. 
The averaged CER of PO3 was better than that of BO2 while retaining the decoding speed. 
We performed the MAPSSWE significance test~\cite{mapsswe}, and the differences of the CERs between BO2 and PO3 under all conditions were statistically significant, $p < 0.001$. The relative CER reduction was 5.1\% on average.

The streaming experiment shows that the proposed streaming system PS4 ($\lambda = 0.01$) greatly outperforms the baseline system BS2. 
The relative CER reduction between the integrated system BS2 and the proposed system PS4 was 9.4\%, and the differences under all conditions were also statistically significant ($p < 0.001$) by performing MAPSSWE significance test. 
These results show the effectiveness of our proposed loss term $\mathcal{L}_{\text{RNNT+KD}}$ in exploiting parallel mixture/single-talker speech data for TS-RNNT.

\vspace{-0.3cm}
\section{Conclusion}
\label{ssec:conclusions}
We have proposed a KD framework that allows us to exploit the parallel mixture/single-talker speech data for improving an end-to-end TS-ASR model, called TS-RNNT. 
We train a teacher RNNT with single-speaker speech and transcription. The pretrained model with target single-talker speech input is used to obtain a clearer posterior of the target speaker than is possible with the conventional TS-RNNT. 
Then the posterior and the TS-RNNT output are used for KD loss computation. 
The posterior of the single-talker ASR with the target speaker's speech input guided the TS-RNNT output to behave similarly, resulting in improved TS-ASR performance. 
Our proposed system TS-RNNT+KD offers better performance than TSE+RNNT and TS-RNNT in both streaming and offline settings. 
The improvement is particularly significant in streaming operation.
\pagebreak

\bibliographystyle{IEEEtran}
\bibliography{mybib,refs}
\end{document}